\documentclass[twocolumn,showpacs,preprintnumbers,prd,superscriptaddress,
nofootinbib,amsmath,amssymb]{revtex4-2}

\usepackage{amsmath,amssymb,mathrsfs}
\usepackage{array}
\usepackage[utf8]{inputenc}
\usepackage[colorlinks=true,citecolor=blue,linkcolor=blue,urlcolor=blue,
breaklinks=true]{hyperref}

\newcommand{\pr}{\partial}
\newcommand{\Shi}{\mathrm{Shi}}
\newcommand{\SBH}{S_{\rm BH}}

\usepackage{amsfonts}
\usepackage[usenames,dvipsnames,svgnames]{xcolor}  
\usepackage{hyperref}   
\definecolor{oxfordblue}{rgb}{0.0, 0.13, 0.28}
\definecolor{burgundy}{rgb}{0.5, 0.0, 0.13}
\definecolor{darkolivegreen}{rgb}{0.33, 0.42, 0.18}
\definecolor{darkblue}{rgb}{0,0,0.5}
\definecolor{richcarmine}{rgb}{0.84, 0.0, 0.25}
\definecolor{darkblue}{rgb}{0,0,0.5}
\definecolor{bluer}{rgb}{0.00,0.50,0.75}{}
\hypersetup{colorlinks=true, citecolor=red, linkcolor=blue,
 urlcolor = magenta, filecolor=magenta}

\begin{document}

\title{Reconstructing $f(R)$ gravity from generalized entropies: exact 
Lagrangians}

\author{Ankit Anand}
\email{anand@iitk.ac.in}
\affiliation{Department of Physics, Indian Institute of Technology, Kanpur 
208016, India}
                
\author{Sahil Devdutt}
\email{devduttsahil@gmail.com}
\affiliation{Department of Physics \& Astronomical Science, Central University 
of Himachal Pradesh, Dharamshala 176215, India}

\author{Kimet Jusufi}
\email{kimet.jusufi@unite.edu.mk}
\affiliation{Physics Department, State University of Tetovo, Ilinden Street nn, 
1200 Tetovo, North Macedonia}

\author{Ayan Chatterjee}
\email{ayan.theory@gmail.com}
\affiliation{Department of Physics \& Astronomical Science, Central University 
of Himachal Pradesh, Dharamshala 176215, India}

 \author{Emmanuel N. Saridakis}
\email{msaridak@noa.gr}
 \affiliation{Institute for Astronomy, Astrophysics, Space Applications and 
Remote Sensing, National Observatory of Athens, 15236 Penteli, Greece}
\affiliation{Departamento de Matem\'{a}ticas, Universidad Cat\'{o}lica del 
	Norte, Avda. Angamos 0610, Casilla 1280, Antofagasta, Chile}
\affiliation{CAS Key Laboratory for Research in Galaxies and Cosmology, 
	School 
	of Astronomy and Space Science,
	University of Science and Technology of China, Hefei 230026, China}	

\begin{abstract}  
Generalized horizon entropies are widely used as  theoretical
modifications of the Bekenstein-Hawking area law, but through the Wald
construction they may also encode modifications of the underlying
gravitational dynamics. We reconstruct metric $f(R)$ gravity from prescribed
entropy-area relations and show that the procedure is intrinsically branch
dependent through the required area-curvature map. On the maximally symmetric
branch, where $A=48\pi/R$ exactly, the reconstruction reduces to a single
quadrature and can be performed non-perturbatively. We obtain closed-form
Lagrangians for several generalized entropies and show that an entropy term
$a\SBH^q$ generates a curvature term proportional to $R^{2-q}$. In
particular, Kaniadakis entropy produces a $1/R$ correction, while logarithmic
entropy generates an $R^2\ln R$ term. We further derive a branch-independent
criterion,
$\pr_R^2f=(ds/dR)d(S/s)/ds$, relating Dolgov-Kawasaki stability directly to
the entropy functional, together with
$m_{\rm sc}^2=S'(s)/(3f_{RR})$ on the maximally symmetric branch.
Comparison with the fixed-mass Schwarzschild-de Sitter branch reveals
different reconstructed Lagrangians and reversed stability properties.
Finally, the weak-isolated-horizon boost charge reproduces the original
generalized entropy. These results establish a direct non-perturbative link
between generalized horizon thermodynamics and modified gravitational
dynamics.

\end{abstract}

\maketitle
 
\section{Introduction}
\label{sec:introduction}

The Bekenstein-Hawking entropy $\SBH=A/(4G)$ provides one of the most
fundamental connections between gravitation, thermodynamics and quantum
theory \cite{Bekenstein1973,Hawking1975}. Nevertheless, there are several
independent reasons why the entropy associated with a gravitational horizon
may depart from the standard area law. Long-range interactions, non-extensive
statistical behavior, relativistic generalizations of statistical mechanics,
non-equilibrium effects, quantum correlations and possible modifications of
the microscopic horizon structure may all lead to generalized entropy
functionals.  In particular, Tsallis  
\cite{Tsallis:1987eu,Tsallis:2012js} and R\'enyi  
\cite{Renyi1961,Majhi:2017zao} entropies 
arise from generalized statistical
frameworks, Kaniadakis entropy 
\cite{Kaniadakis:2001tcv,Kaniadakis:2002zz}  is motivated by 
a relativistic extension of
statistical mechanics, while Barrow entropy \cite{Barrow:2020tzx} is associated 
with a possible
fractal deformation of the horizon induced by quantum-gravitational effects.
Despite their different motivations, such entropies can be applied to
gravitational horizons and typically recover the Bekenstein-Hawking form in
 appropriate limits
\cite{Czinner:2015eyk,Biro:2013cra,Saridakis:2020zol,Kaniadakis:2005zk}. 
Additionally, when applied to cosmological horizons, they can lead to modified 
cosmological scenarios through the gravity-thermodynamics conjecture 
\cite{Lymperis:2018iuz,Barboza:2014yfe,
Abreu:2012msk,Lima:2001vq,Abreu:2017hiy,Luciano:2021mto,
Luciano:2025hjn,
Headrick:2010zt,Hung:2011nu,Chen:2013kpa,  
 Abreu:2018phv, Hernandez-Almada:2021rjs, 
Luciano:2024bco, Luciano:2022eio,Luciano:2022knb,  
  Mamon:2020spa,
Srivastava:2020cyk,Figliolia:2026sma}, or to extended 
holographic dark energy scenarios
\cite{Saridakis:2018unr, Zadeh:2018poj,Ghaffari:2018wks,
Aditya:2019bbk,Huang:2019hex,Pandey:2021fvr, 
 Luciano:2025elo, 
 Maity:2019qbv,
Iqbal:2019ooy,  
Drepanou:2021jiv,Hernandez-Almada:2021aiw, 
Sadeghi:2022fow,
Gonzalez-Espinoza:2026iyh,Bhupinder:2026rdt, 
 Anagnostopoulos:2020ctz, 
Nojiri:2021jxf,Oliveros:2022biu,Huang:2021zgj,Yarahmadi:2024oqv,Paul:2022doh,
Yarahmadi:2024afr,
Luciano:2022ffn,Luciano:2022hhy,Luciano:2023wtx, 
Basilakos:2023seo}.

Since the horizon entropy is
modified, it is natural to ask what gravitational dynamics correspond to
this modification. For any diffeomorphism-invariant theory the entropy of a
stationary horizon is determined by the associated Noether charge
\cite{Wald:1993nt,Iyer:1994ys,Jacobson:1993vj}. In metric $f(R)$ gravity 
\cite{Nojiri:2010wj,CANTATA:2021asi} the
Wald entropy is
$
S_{\rm Wald}=\frac{A}{4G}f_R $.
Thus, a prescribed entropy-area relation determines $f_R$ on the horizon and
can in principle be used to reconstruct the gravitational Lagrangian. This
inverse problem is also motivated by the broader relation between
gravitational dynamics and horizon thermodynamics suggested by
thermodynamic and emergent-gravity approaches
\cite{Wald:1993nt,Iyer:1994ys,Jacobson:1993vj} (see also 
\cite{Anand:2025rjg,Jusufi:2025rlr,Anand:2025cer}).

The main difficulty is that the entropy is naturally a function of the
horizon area, whereas the Lagrangian is a function of the curvature. Hence,
an area-curvature relation $A(R)$ must be specified. As we show, this relation
is not unique. Schwarzschild-de Sitter constant-curvature geometries depend
on both the curvature and the mass parameter, and therefore the horizon area
is not uniquely fixed by $R$. The reconstruction becomes well defined only
after a one-parameter branch in this space of geometries is selected.
Different branches consequently lead to different entropy-Lagrangian
correspondences.

We investigate two physically distinct choices. For a Schwarzschild-de
Sitter black hole of fixed mass, the area-curvature relation is known through
an expansion around the Schwarzschild solution. On the maximally symmetric
branch, however, the cosmological horizon satisfies
$
A=\frac{48\pi}{R} $,
which is exact and mass independent. In this case the Wald relation reduces
to a single quadrature and the corresponding $f(R)$ Lagrangian can be
reconstructed without expanding in the entropy deformation parameter.

Using this exact construction, we obtain closed-form Lagrangians for the
Bekenstein-Hawking, Tsallis-Cirto, Barrow, R\'enyi, Kaniadakis, logarithmic
and power-law entanglement entropies. We further derive a general mapping
between entropy and curvature powers. An entropy contribution proportional
to $\SBH^q$ generates a curvature contribution proportional to $R^{2-q}$,
with $q=2$ yielding a logarithmic term. Hence, the exponent is reflected
about the extensive value $q=1$, directly determining the infrared or
higher-curvature character of the induced modification. In particular, the
Kaniadakis entropy generates at leading order the $1/R$ modification of
Carroll, Duvvuri, Trodden and Turner \cite{Carroll:2003wy}, while the
logarithmically corrected entropy produces the $R^2\ln R$ structure appearing
in one-loop effective descriptions
\cite{Kaul:2000kf,Carlip:2000nv,Solodukhin:2011gn}.

Furthemore, we derive general viability conditions directly from the entropy
functional. A branch-independent relation expresses $f_{RR}$ through the
monotonicity of $S(s)/s$, where $s=\SBH$, and therefore turns the
Dolgov-Kawasaki condition into an entropic criterion whose sign is reversed
between the two branches. On the maximally symmetric branch the scalaron
mass can additionally be written directly in terms of $S'(s)$. Hence, ghost
freedom, Dolgov-Kawasaki stability and the absence of a tachyonic scalaron
can all be related to properties of the entropy itself.

For comparison, we analyze the massive branch and reconstruct the R\'enyi,
Barrow and Kaniadakis models within its regime of validity, examining their
constant-curvature vacua and scalaron stability. The comparison shows that
the two branches are not different approximations to the same theory, but
different reconstructions associated with different horizon families.
Finally, we express the reconstructed theories in first-order scalar-tensor
form and study them within the covariant phase space of weak isolated
horizons
\cite{Ashtekar:1998sp,Ashtekar:2000hw,Ashtekar:2001is,Ashtekar:2004cn}.
The corresponding conserved boost charge reproduces the generalized entropy,
providing an independent quasi-local consistency check.

The structure of the paper is as follows. In section~\ref{sec:framework} we
present the reconstruction framework, the area-curvature ambiguity, the two
branches and the exact maximally symmetric construction. In
section~\ref{sec:exactproperties} we derive the exact Lagrangians and examine
their constant-curvature vacua, stability and infrared properties. In
section~\ref{sec:massive} we study the massive branch, compare the two
reconstructions and derive the corresponding weak isolated-horizon charges.
Finally, in section~\ref{sec:conclusions} we summarize the results and discuss
possible extensions.
We use $G=c=\hbar=1$ except where $G$ is displayed for clarity, and write
$\pr_Rf\equiv f_R$ and $\pr_R^2f\equiv f_{RR}$.

\section{Reconstruction framework}
\label{sec:framework}

In this section we formulate the reconstruction procedure that relates a
prescribed horizon entropy to an $f(R)$ gravitational Lagrangian. We first
identify the role of the area-curvature relation and its branch dependence,
and then derive the corresponding viability conditions and the exact
reconstruction on the maximally symmetric branch.

\subsection{Master relation and area-curvature ambiguity}
\label{subsec:master}

For the $f(R)$ action \cite{Nojiri:2010wj, Starobinsky:2007hu, Nojiri:2006gh, 
Bertolami:2007gv, Song:2006ej, Nojiri:2007as, Lobo:2009ip, Tsujikawa:2007xu, 
Nojiri:2007cq, Faulkner:2006ub, Chiba:2006jp}
\begin{equation}
I=\frac{1}{16\pi G}\int d^4x\,\sqrt{-g}\,f(R) \ ,
\label{eq:action}
\end{equation}
the Wald entropy of a stationary horizon is \cite{Nojiri:2010wj,CANTATA:2021asi}
\begin{equation}
S_{\rm Wald}=\frac{A}{4G}\,\pr_R f \ ,
\label{eq:wald}
\end{equation}
which reduces to $A/4G$ for $f(R)=R-2\Lambda$. Demanding that
Eq.~\eqref{eq:wald} is equal to a prescribed generalized entropy $S(A)$ gives
the master relation of the reconstruction programme
\begin{equation}
\pr_R f=\frac{4G\,S(A)}{A} \ .
\label{eq:master}
\end{equation}
Since most generalized entropies recover $\SBH$ as their corresponding
deformation parameter vanishes, one may write the generalized entropy as an
expansion in this deformation parameter according to
\begin{equation}
S=\SBH\left[1+\epsilon\,\Phi(A)+\mathcal{O}(\epsilon^2)\right] \ ,
\label{eq:entropyexpansion}
\end{equation}
where $\epsilon$ is the deformation parameter and $\Phi(A)$ is a
model-dependent function of the horizon area. In this language, the Einstein
limit is approached continuously and $\epsilon$ acquires the interpretation of
an effective coupling that measures departures from General Relativity. We
shall use Eq.~\eqref{eq:entropyexpansion} in the analysis of the massive branch
below. However, as we will show, the full Lagrangian can also be recovered in
closed form without such a perturbative expansion if one assumes the existence
of a maximally symmetric solution.

We mention here that relation \eqref{eq:master} is exact. It determines $\pr_R 
f$ as a function of
the horizon area, whereas the Lagrangian is a function of the Ricci scalar.
Hence, the reconstruction is complete only once a relation $A=A(R)$ has been
supplied. Let us therefore examine this relation in more detail.

We consider Schwarzschild-de Sitter geometries within metric $f(R)$ gravity,
given by
\begin{equation}
\begin{gathered}
ds^2=-h(r)dt^2+h(r)^{-1}dr^2+r^2d\Omega^2 \ ,\\[2pt]
h(r)=1-\frac{2Gm}{r}-\frac{Rr^2}{12} \ ,
\end{gathered}
\label{eq:sds}
\end{equation}
whose horizon radius satisfies
\begin{equation}
\frac{R}{12}r_h^3-r_h+2Gm=0 \ .
\label{eq:cubic}
\end{equation}
The viability conditions for the existence of such solutions will be discussed
in the following subsection. Since Eq.~\eqref{eq:cubic} may be solved for the 
mass as
$2Gm=r_h-Rr_h^3/12$, for any chosen $R$ and $r_h$ it fixes a
corresponding mass parameter $m$. Thus, at the level of the
Schwarzschild-de Sitter ansatz, the horizon area is not uniquely determined
by the scalar curvature only. Consequently, there is no unique canonical function
$A(R)$, and Eq.~\eqref{eq:master} becomes well posed only after a
one-parameter curve through this family specified by $(M,\Lambda)$ has been selected. Different curves 
define
different reconstructions, and this choice is the principal input of the
method. We stress that this is not an approximation. Two natural choices are
the massive branch and the maximally symmetric branch.

\paragraph*{(i) The massive branch.}
The horizon radius can be found by solving Eq.~\eqref{eq:cubic} for $r_h$.
Expanding around the Schwarzschild point $R=0$, one obtains
$r_h=2GM(1+G^2M^2R/3)+\mathcal{O}(R^2)$. Therefore, the horizon area can be
approximated as
\begin{equation}
A(R)=A_0(1+cR),    \ \ \  A_0\equiv16\pi G^2M^2,   \ \ \ 
c\equiv\frac{2G^2M^2}{3} \ .
\label{eq:map_massive}
\end{equation}
This map underpins the existing reconstruction method and will be used in the
analysis of the massive branch below. It is   the origin of the constant
$c$ appearing there and it has two limitations. It retains a dependence on 
$M$, 
and hence the
reconstructed Lagrangian is not universal. Moreover, it is accurate only to
leading order in the curvature expansion around $R=0$, and therefore any
reconstruction based on this map is controlled within this curvature regime.

\paragraph*{(ii) The maximally symmetric branch.}
Setting $m=0$ instead leaves a one-parameter family of de Sitter geometries
with $R=12/L^2$ and a cosmological horizon of radius $L$, yielding
\begin{equation}
A(R)=\frac{48\pi}{R} \ .
\label{eq:map_ds}
\end{equation}
Equation~\eqref{eq:map_ds} is exact, involves no expansion, and carries no free
mass parameter. It therefore defines a reconstruction that is exact in the
deformation parameter, and it is this property that allows for the closed-form
Lagrangians derived below.

Both branches are legitimate, and neither is preferred on thermodynamic
grounds alone. Each imposes Eq.~\eqref{eq:master} along a different curve of
vacua and each answers a different physical question, namely what Lagrangian
reproduces a given entropy for a black hole of fixed mass or for a cosmological
horizon. They agree, as they must, when the deformation is switched off, with
both returning $f(R)=R-2\Lambda$. However, they differ in the structure of the
induced corrections, as we will discuss in detail below.

\subsection{Viability conditions}
\label{subsec:viability}

A reconstructed Lagrangian is admissible only if it supports a maximally
symmetric vacuum and a healthy scalar sector. Varying Eq.~\eqref{eq:action}
gives
\begin{equation}
f_RR_{\mu\nu}-\tfrac12fg_{\mu\nu}
+(g_{\mu\nu}\Box-\nabla_\mu\nabla_\nu)f_R=0 \ ,
\label{eq:fieldeq}
\end{equation}
whose trace, $Rf_R-2f+3\Box f_R=0$, reduces at constant curvature to the
algebraic condition
\begin{equation}
R_0f_R(R_0)-2f(R_0)=0 \ .
\label{eq:cc}
\end{equation}
Every real root of  equation \eqref{eq:cc} supplies a Schwarzschild-(anti-)de 
Sitter
background of the form \eqref{eq:sds}. Depending on the sign of the root,
expression \eqref{eq:sds} with $R\to R_0$ describes a Schwarzschild-de Sitter
($R_0>0$), Schwarzschild ($R_0=0$), or Schwarzschild-anti-de Sitter
($R_0<0$) geometry. Hence, the reconstructed theory possesses genuine
black-hole solutions rather than only a maximally symmetric vacuum.

Absence of ghosts requires a positive effective Newton constant,
$f_R(R_0)>0$. Absence of the Dolgov-Kawasaki instability requires
$f_{RR}(R_0)>0$ \cite{Dolgov:2003px,Faraoni:2006sy}. Additionally, the
scalaron, namely the additional scalar mode of metric $f(R)$ gravity, must be
non-tachyonic,
\begin{equation}
m_{\rm sc}^2=
\frac{f_R(R_0)-R_0f_{RR}(R_0)}
{3f_{RR}(R_0)}>0 \ .
\label{eq:scalaron}
\end{equation}
These are the standard viability criteria
\cite{DeFelice:2010aj,Sebastiani:2010kv,Multamaki:2006zb}, and we apply
them to every reconstruction below.

\subsection{Exact reconstruction on the maximally symmetric branch}
\label{sec:exact}

On the branch \eqref{eq:map_ds} the Bekenstein-Hawking entropy of the
cosmological horizon is
\begin{equation}
s\equiv\SBH=\frac{A}{4G}=\frac{\sigma}{R} \ , \qquad
\sigma\equiv\frac{12\pi}{G} \ ,
\label{eq:sofR}
\end{equation}
so that $R=\sigma/s$ and $dR=-\sigma s^{-2}ds$. Writing the generalized entropy
as a function of $s$, Eq.~\eqref{eq:master} reads $\pr_Rf=S(s)/s$, and
integrating gives the exact master formula
\begin{equation}
f(R)=-\sigma\int^{\,\sigma/R}\frac{S(s')}{s'^{\,3}}\,ds'+c_1 \ ,
\label{eq:exactmaster}
\end{equation}
with $c_1$ fixed by the undeformed limit. Equation~\eqref{eq:exactmaster} is the
central result of this paper. It contains no expansion of any kind. Given
$S(s)$, a single elementary quadrature returns the Lagrangian.

Its consistency can be directly verified. Differentiating, one obtains
\begin{equation}
\pr_Rf=-\sigma\frac{S(s)}{s^3}\frac{ds}{dR}
=\frac{\sigma^2}{R^2}\frac{S(s)}{s^3}
=\frac{S(s)}{s} \ ,
\end{equation}
which reproduces Eq.~\eqref{eq:master} identically. Moreover, for the undeformed
entropy $S(s)=s$,
\begin{equation}
f(R)=-\sigma\int^{\sigma/R}\frac{ds'}{s'^{\,2}}+c_1
=\frac{\sigma}{s}+c_1=R-2\Lambda \ ,
\end{equation}
and thus Einstein gravity is recovered identically rather than asymptotically.
This is a stronger statement than can be made on the massive branch, where the
Einstein limit holds only to the order at which the map
\eqref{eq:map_massive} is valid.

The exact master formula also allows for a general relation between the
functional form of the entropy and that of the resulting Lagrangian. Suppose
that the entropy admits a generalized power-series representation
\begin{equation}
S(s)=\sum_k a_k\,s^{q_k} \ ,
\label{eq:series}
\end{equation}
with arbitrary real exponents $q_k$. Term-by-term integration of
Eq.~\eqref{eq:exactmaster} gives
\begin{equation}
f(R)=\sum_k\frac{a_k\sigma^{\,q_k-1}}{2-q_k}\,R^{\,2-q_k}+c_1 \ ,
\qquad q_k\neq2 \ ,
\label{eq:lemma}
\end{equation}
while the exceptional exponent $q_k=2$ contributes $a_k\sigma\ln R$ in place
of the corresponding power. Equation~\eqref{eq:lemma} may therefore be
summarized through the following rule.

\noindent\emph{A term $a\,\SBH^{\,q}$ in the entropy generates a term
$\propto R^{\,2-q}$ in the Lagrangian.}

\noindent
Hence, the exponent is reflected about $q=1$. The extensive term $q=1$ maps
to the Einstein term $R$.
More importantly, entropy contributions with $q>1$ generate infrared
curvature corrections with $2-q<1$. This
inversion is the characteristic feature of the maximally symmetric branch.
On the massive branch, where $s\propto1+cR$, the exponents are preserved,
while here they are reflected. This follows directly from the inverse relation
\eqref{eq:sofR} between entropy and curvature on a de Sitter horizon, where a
larger horizon corresponds to a smaller curvature.

\section{Exact Lagrangians and physical properties}
\label{sec:exactproperties}

In this section we apply the exact reconstruction derived above to the
generalized entropies of interest. We obtain the corresponding closed-form
$f(R)$ Lagrangians and then investigate their constant-curvature vacua,
stability conditions and infrared or higher-curvature behavior.

\subsection{Exact Lagrangians}
\label{subsec:exactresults}

We now apply Eq.~\eqref{eq:exactmaster} to the entropies of interest. All
results below have been verified by differentiation against
Eq.~\eqref{eq:master}, both symbolically and by numerical quadrature, and are
collected in Table~\ref{tab:exact}.

\paragraph*{Tsallis-Cirto 
\cite{Tsallis:1987eu,Tsallis:2012js} and  Barrow  \cite{Barrow:2020tzx}
entropies.}
For $S_T=\gamma s^{\delta}$ the relation \eqref{eq:lemma} applies with a
single term,
\begin{equation}
f_T(R)=\frac{\gamma\sigma^{\,\delta-1}}{2-\delta}\,R^{\,2-\delta}
-2\Lambda \ , \qquad \delta\neq2 \ .
\label{eq:tsallis}
\end{equation}
The Barrow entropy is the case $\gamma=1$ and $\delta=1+\Delta/2$, giving the
pure power law
\begin{equation}
f_B(R)=\frac{\sigma^{\Delta/2}}{1-\Delta/2}\,R^{\,1-\Delta/2}-2\Lambda \ .
\label{eq:barrowexact}
\end{equation}
The fractal exponent appears directly as a shift of the power of the Ricci
scalar, and Eq.~\eqref{eq:barrowexact} is exact for all
$\Delta\in[0,1]$, without any smallness assumption. The marginal case
$\delta=2$ gives instead a purely logarithmic Lagrangian,
\begin{equation}
f_T(R)\big|_{\delta=2}=\gamma\sigma\ln R-2\Lambda \ .
\label{eq:tsallislog}
\end{equation}

\paragraph*{R\'enyi entropy \cite{Renyi1961,Majhi:2017zao}.}
For $S(s)=\lambda^{-1}\ln(1+\lambda s)$ the integral in
Eq.~\eqref{eq:exactmaster} is elementary,
\begin{align}
f_{\mathcal R}(R)=\;&
\frac{R^{2}}{2\lambda\sigma}\ln\!\left(1+\frac{\lambda\sigma}{R}\right)
+\frac{R}{2}\notag\\
&-\frac{\lambda\sigma}{2}
\ln\!\left(\frac{R+\lambda\sigma}{\sigma}\right)-2\Lambda \ .
\label{eq:renyiexact}
\end{align}
It is worth noting that the R\'enyi reconstruction, which on the massive
branch requires a dilogarithm, is elementary here. Expanding for small
$\lambda$, one obtains
\begin{equation}
f_{\mathcal R}(R)=R-\frac{\lambda\sigma}{2}\ln R-2\Lambda
+{\rm const}+\mathcal{O}(\lambda^{2}) \ ,
\label{eq:renyiexpand}
\end{equation}
and hence the leading correction is a curvature logarithm.

\paragraph*{Kaniadakis entropy \cite{Kaniadakis:2001tcv,Kaniadakis:2002zz}.}
For $S(s)=\kappa^{-1}\sinh(\kappa s)$, two integrations by parts give
\begin{align}
f_K(R)=\;&
\frac{R^{2}}{2\kappa\sigma}\sinh\!\frac{\kappa\sigma}{R}
+\frac{R}{2}\cosh\!\frac{\kappa\sigma}{R}\notag\\
&-\frac{\kappa\sigma}{2}\,
\Shi\!\left(\frac{\kappa\sigma}{R}\right)-2\Lambda \ ,
\label{eq:kaniexact}
\end{align}
where $\Shi(x)=\int_0^x\sinh(t)\,t^{-1}dt$ is the hyperbolic sine integral.
Expanding for small $\kappa$, we find
\begin{equation}
f_K(R)=R-\frac{\kappa^{2}\sigma^{2}}{6R}-2\Lambda
+\mathcal{O}(\kappa^{4}) \ .
\label{eq:kaniexpand}
\end{equation}
The leading Kaniadakis correction on the de Sitter branch is precisely the
$1/R$ modification proposed by Carroll, Duvvuri, Trodden and Turner as a
geometric origin of late-time acceleration \cite{Carroll:2003wy}, with the
inverse-curvature scale fixed by the deformation parameter as
$\kappa^2\sigma^2/6$. Its stability will be examined below.

\paragraph*{Logarithmically corrected entropy.}
For $S(s)=s+\alpha\ln s$, characteristic of loop quantum gravity and
microstate counting \cite{Kaul:2000kf,Carlip:2000nv}, one obtains
\begin{equation}
f_{\log}(R)=R+\frac{\alpha R^{2}}{4\sigma}
\left[1+2\ln\frac{\sigma}{R}\right]-2\Lambda \ .
\label{eq:logexact}
\end{equation}
The correction has the $R^2\ln R$ form generated by the conformal anomaly and
by one-loop effective actions in curved space \cite{Solodukhin:2011gn}.
Hence, a logarithmic entropy correction leads directly to a logarithmic
curvature form factor, and this relation holds exactly in $\alpha$ rather than
perturbatively.

\paragraph*{Power-law entanglement entropy.}
For $S(s)=s(1-\mu s^{\beta})$ \cite{Das:2007mj}, Eq.~\eqref{eq:lemma} gives
\begin{equation}
f_{\rm pl}(R)=R-\frac{\mu\sigma^{\beta}}{1-\beta}\,R^{\,1-\beta}
-2\Lambda \ , \qquad \beta\neq1 \ .
\label{eq:powerlaw}
\end{equation}

\begin{table*}[t]
\centering
\begin{ruledtabular}
\begin{tabular}{lcccc}
Entropy & $S(s)$, $\ s\equiv\SBH=\sigma/R$ & Exact $f(R)+2\Lambda$ &
Leading correction & DK stability \\
\hline
Bekenstein-Hawking & $s$ & $R$ & - & marginal \\[3pt]
Tsallis-Cirto & $\gamma s^{\delta}$ &
$\dfrac{\gamma\sigma^{\delta-1}}{2-\delta}R^{2-\delta}$ &
$\propto R^{2-\delta}$ & only if $\delta<1$ \\[8pt]
Barrow & $s^{1+\Delta/2}$ &
$\dfrac{\sigma^{\Delta/2}}{1-\Delta/2}R^{1-\Delta/2}$ &
$\propto R^{1-\Delta/2}$ & unstable ($\Delta>0$) \\[8pt]
R\'enyi & $\lambda^{-1}\ln(1+\lambda s)$ &
Eq.~\eqref{eq:renyiexact}, elementary &
$-\tfrac{\lambda\sigma}{2}\ln R$ & stable, all $\lambda>0$ \\[3pt]
Kaniadakis & $\kappa^{-1}\sinh(\kappa s)$ &
Eq.~\eqref{eq:kaniexact}, hyperbolic sine integral &
$-\tfrac{\kappa^2\sigma^2}{6}R^{-1}$ & unstable \\[3pt]
Logarithmic & $s+\alpha\ln s$ &
Eq.~\eqref{eq:logexact} & $\propto R^{2}\ln R$ &
$\alpha(1-\ln s)<0$ \\[3pt]
Power-law entanglement & $s(1-\mu s^{\beta})$ &
$R-\dfrac{\mu\sigma^{\beta}}{1-\beta}R^{1-\beta}$ &
$\propto R^{1-\beta}$ & stable if $\mu\beta>0$ \\
\end{tabular}
\end{ruledtabular}
\caption{Exact $f(R)$ Lagrangians reconstructed from generalized horizon
entropies on the maximally symmetric branch through the master formula
\eqref{eq:exactmaster}, with $\sigma=12\pi/G$. No expansion in the deformation
parameter is involved. All entries have been verified by differentiation
against Eq.~\eqref{eq:master} and by numerical quadrature. The last column
follows from the sub-extensivity criterion \eqref{eq:criterion}.}
\label{tab:exact}
\end{table*}

\subsection{Constant-curvature vacua}
\label{subsec:exactvacua}

The reconstructed Lagrangians are physically admissible only if they possess
maximally symmetric vacua. On this branch the trace condition
\eqref{eq:cc} can be written in a general form. Setting
$f(R)=F(s)-2\Lambda$, with
$F(s)\equiv-\sigma\int^{s}S(s')s'^{-3}ds'$, and using
$\pr_Rf=S(s)/s$ and $R=\sigma/s$, Eq.~\eqref{eq:cc} becomes
\begin{equation}
\frac{\sigma\,S(s_0)}{s_0^{2}}-2F(s_0)+4\Lambda=0 \ ,
\label{eq:vacgeneral}
\end{equation}
which is an algebraic equation for the horizon entropy
$s_0=\sigma/R_0$ of the vacuum. Equation~\eqref{eq:vacgeneral} is exact
and requires only the entropy functional and its quadrature. Applied to the
single-power family \eqref{eq:tsallis}, it gives
\begin{equation}
\frac{\gamma\delta\sigma^{\delta-1}}{2-\delta}
R_0^{\,2-\delta}=4\Lambda \ ,
\end{equation}
and thus a real vacuum exists whenever
$\Lambda(2-\delta)/\delta>0$, with
\begin{equation}
R_0=\left[
\frac{4\Lambda(2-\delta)}
{\gamma\delta\sigma^{\delta-1}}
\right]^{\frac{1}{2-\delta}} \ .
\label{eq:R0exact}
\end{equation}
Setting $\gamma=\delta=1$ returns $R_0=4\Lambda$, namely the Einstein
value, as required. For the Barrow entropy
$\delta=1+\Delta/2<2$ throughout the physical range
$\Delta\in[0,1]$, and hence a real de Sitter vacuum exists for every
admissible Barrow exponent. The same conclusion holds for the power-law 
entanglement entropy for
$0<\beta<1$ and $\Lambda>0$, for which the constant-curvature equation
admits at least one positive root.

For the R\'enyi and Kaniadakis reconstructions
Eq.~\eqref{eq:vacgeneral} is transcendental and must be solved numerically.
Nevertheless, the existence of a root near the Einstein value follows from
the implicit function theorem. In the zero-deformation limit the left-hand
side of Eq.~\eqref{eq:vacgeneral} reduces to
$-\sigma/s_0+4\Lambda$, which vanishes at
$s_0=\sigma/(4\Lambda)$ and has a non-vanishing derivative there.
Therefore, a root continuously connected to the Einstein solution exists
for sufficiently small deformation.

Ghost freedom follows directly. Since $\pr_Rf=S(s)/s$ and both $S$ and
$s$ are positive, the effective Newton constant is automatically positive
on this branch,
\begin{equation}
\pr_Rf=\frac{S(s)}{s}>0 \ ,
\end{equation}
for every entropy in Table~\ref{tab:exact}. 
Hence, the reconstruction does not introduce ghosts. The remaining viability
conditions concern the sign of $f_{RR}$ and the scalaron mass, which determine
the Dolgov-Kawasaki and tachyonic stability, respectively. These conditions
will be examined below.

\subsection{Extensivity, stability and infrared structure}
\label{sec:stability}
\label{subsec:lemma}

The exact reconstruction allows for a general statement about stability which
holds for both branches. Differentiating the master relation
\eqref{eq:master}, written as $\pr_Rf=S(s)/s$, and using the chain rule, we
obtain
\begin{equation}
\pr^2_Rf=\frac{ds}{dR}\,
\frac{d}{ds}\!\left[\frac{S(s)}{s}\right] \ .
\label{eq:lemma_stab}
\end{equation}
Equation~\eqref{eq:lemma_stab} makes no reference to a specific branch, since
the entire branch dependence lies in the sign of $ds/dR$. The
Dolgov-Kawasaki condition $\pr_R^2f>0$ is therefore determined by the
monotonicity of $S(s)/s$, namely by whether the generalized entropy is sub-
or super-extensive, with the branch determining which behavior is required.

\paragraph*{Maximally symmetric branch.}
In this case $s=\sigma/R$, and hence $ds/dR=-\sigma/R^2<0$. Therefore,
\begin{equation}
\frac{d}{ds}\!\left[\frac{S(s)}{s}\right]<0
\quad\Longleftrightarrow\quad
\pr^2_Rf>0 \ .
\label{eq:criterion}
\end{equation}
Thus, on the de Sitter branch the reconstructed theory is free of the
Dolgov-Kawasaki instability if and only if the generalized entropy is
sub-extensive.

On this branch the scalaron condition can also be expressed directly in
terms of the entropy functional. Since $s=\sigma/R$, one has
\begin{equation}
R\pr_R^2f
=-s\frac{d}{ds}\!\left[\frac{S(s)}{s}\right] \ ,
\end{equation}
and therefore
\begin{equation}
\pr_Rf-R\pr_R^2f
=
\frac{S(s)}{s}
+s\frac{d}{ds}\!\left[\frac{S(s)}{s}\right]
=
\frac{dS}{ds} \ .
\end{equation}
Using Eq.~\eqref{eq:scalaron}, we thus obtain
\begin{equation}
m_{\rm sc}^2=
\frac{S'(s)}{3\pr_R^2f} \ .
\label{eq:scalaronentropy}
\end{equation}
Hence, whenever the entropy is monotonically increasing, $S'(s)>0$, the
Dolgov-Kawasaki condition $\pr_R^2f>0$ automatically implies
$m_{\rm sc}^2>0$. Together with $\pr_Rf=S(s)/s>0$, the three viability
requirements can therefore be expressed directly in terms of the entropy
functional on the maximally symmetric branch.

\paragraph*{Massive branch.}
Here $s=s_0(1+cR)$, with $s_0=A_0/4G$ and $c>0$, and therefore
$ds/dR>0$. Hence, the inequality is reversed and Dolgov-Kawasaki stability
requires super-extensivity, namely $d(S/s)/ds>0$.

Let us now apply these criteria to the generalized entropies considered above
on the maximally symmetric branch.

\paragraph*{R\'enyi entropy.}
For $S/s=\ln(1+\lambda s)/(\lambda s)$, one obtains
\begin{equation}
\frac{d}{ds}\!\left[\frac{S}{s}\right]
=\frac{1}{\lambda s^{2}}
\left[\frac{\lambda s}{1+\lambda s}-\ln(1+\lambda s)\right]<0 \ ,
\label{eq:renyimono}
\end{equation}
where the inequality follows from $x/(1+x)<\ln(1+x)$ for $x>0$.
Therefore, the R\'enyi reconstruction is Dolgov-Kawasaki stable for every
$\lambda>0$, without any further restriction on the deformation parameter.
Moreover,
\begin{equation}
S'(s)=\frac{1}{1+\lambda s}>0 \ ,
\end{equation}
and hence Eq.~\eqref{eq:scalaronentropy} shows that the scalaron is
non-tachyonic as well.

\paragraph*{Tsallis-Cirto and Barrow entropies.}
In this case $S/s=\gamma s^{\delta-1}$ and
\begin{equation}
\frac{d}{ds}\!\left[\frac{S}{s}\right]
=\gamma(\delta-1)s^{\delta-2} \ ,
\end{equation}
which is positive for $\delta>1$. Hence, the reconstructed theory is unstable
for every Barrow exponent $\Delta>0$, while the sub-extensive Tsallis-Cirto
regime $\delta<1$ is Dolgov-Kawasaki stable. For $\gamma>0$ and
$\delta>0$ one also has $S'(s)=\gamma\delta s^{\delta-1}>0$. Therefore,
the stable Tsallis-Cirto regime has a non-tachyonic scalaron, while the
Barrow reconstruction has $m_{\rm sc}^2<0$ whenever $\Delta>0$.

\paragraph*{Kaniadakis entropy.}
For $S/s=\sinh(\kappa s)/(\kappa s)$, we obtain
\begin{equation}
\frac{d}{ds}\!\left[\frac{S}{s}\right]
=\frac{\kappa s\cosh(\kappa s)-\sinh(\kappa s)}
{\kappa s^{2}}>0 \ ,
\end{equation}
for all $\kappa\neq0$. Thus, the Kaniadakis reconstruction is
Dolgov-Kawasaki unstable throughout. Since
\begin{equation}
S'(s)=\cosh(\kappa s)>0 \ ,
\end{equation}
Eq.~\eqref{eq:scalaronentropy} further implies $m_{\rm sc}^2<0$ on this
branch.

\paragraph*{Logarithmic and power-law entanglement entropies.}
For $S=s+\alpha\ln s$, one has
$S/s=1+\alpha s^{-1}\ln s$ and
\begin{equation}
\frac{d}{ds}\!\left[\frac{S}{s}\right]
=\frac{\alpha(1-\ln s)}{s^{2}} \ .
\end{equation}
Therefore, Dolgov-Kawasaki stability requires
$\alpha(1-\ln s)<0$. A negative logarithmic coefficient is stabilizing for
horizons with $s<e$ and destabilizing for $s>e$, while the opposite behavior
arises for $\alpha>0$. Since realistic horizons have $s\gg1$, the physically
relevant condition is $\alpha>0$. The scalaron condition additionally
requires
\begin{equation}
S'(s)=1+\frac{\alpha}{s}>0 \ .
\end{equation}

For $S=s(1-\mu s^{\beta})$, one finds
\begin{equation}
\frac{d}{ds}\!\left[\frac{S}{s}\right]
=-\mu\beta s^{\beta-1} \ ,
\end{equation}
and therefore Dolgov-Kawasaki stability requires $\mu\beta>0$. In addition,
Eq.~\eqref{eq:scalaronentropy} requires
\begin{equation}
S'(s)=1-\mu(1+\beta)s^\beta>0
\end{equation}
for the scalaron to remain non-tachyonic.

The Kaniadakis case provides an independent check of the above criterion. Its
leading Lagrangian \eqref{eq:kaniexpand} is the $1/R$ model proposed in
\cite{Carroll:2003wy}, with $\mu^4=\kappa^2\sigma^2/6$. This model is known
to suffer from the Dolgov-Kawasaki instability
\cite{Dolgov:2003px,Faraoni:2006sy}. Hence, the instability obtained directly
from the super-extensive character of $\sinh(\kappa s)/(\kappa s)$ reproduces
the known dynamical result without the need to analyze the field equations.

On the massive branch the sign in Eq.~\eqref{eq:lemma_stab} is inverted.
Thus, the Barrow and Kaniadakis reconstructions satisfy the
Dolgov-Kawasaki condition, while the R\'enyi reconstruction does not. The
explicit scalaron analysis below provides an independent check of this
behavior.

The stability criterion is also closely related to the infrared structure of
the exact Lagrangians. According to Eq.~\eqref{eq:lemma}, an entropy term
$a\SBH^q$ generates a term proportional to $R^{2-q}$ in the Lagrangian.
Therefore, entropy contributions with $q>1$ generate curvature terms with
power smaller than unity and hence infrared modifications. However, the
infrared or ultraviolet character is determined by the exponent $q$, whereas
Dolgov-Kawasaki stability is determined by the sign of
$d(S/s)/ds$. These two properties are therefore related but should not be
identified.

 This distinction is important when the coefficient of the entropy correction
is negative. For example, the power-law entanglement correction scales with
an exponent larger than unity and therefore generates the infrared term
$R^{1-\beta}$. Nevertheless, for $\mu\beta>0$ it satisfies
$d(S/s)/ds<0$ and is Dolgov-Kawasaki stable on the maximally symmetric
branch.

The exponent mapping determines where in the curvature regime the
reconstructed theories differ from General Relativity. In particular, the
Barrow Lagrangian $R^{1-\Delta/2}$ has an exponent below unity, and its ratio
to the Einstein term grows without bound as $R\to0$ and tends to zero as
$R\to\infty$. Similarly, the Kaniadakis correction
$-\kappa^2\sigma^2/(6R)$ diverges as $R\to0$, while the leading R\'enyi
correction $-\lambda\sigma\ln R/2$ becomes relevant at small curvature.

On the other hand, the logarithmic entropy correction produces the
$R^2\ln R$ structure of Eq.~\eqref{eq:logexact}, which has an ultraviolet
character. This case should be distinguished from the exceptional power
$q=2$ in Eq.~\eqref{eq:lemma}. A term proportional to $s^2$ in the entropy
generates a logarithmic term $\ln R$ in the Lagrangian, whereas a term
proportional to $\ln s$ in the entropy generates the $R^2\ln R$ structure.

This behavior follows from the inverse area-curvature relation on the
maximally symmetric branch. Since a large de Sitter horizon corresponds to
small curvature according to Eq.~\eqref{eq:sofR}, entropy contributions with
powers larger than the area term are mapped to curvature contributions that
become important in the infrared. Thus, entropy corrections motivated by
quantum-gravitational considerations may appear on this branch as infrared
modifications of the gravitational Lagrangian.

The Kaniadakis case makes this connection particularly clear. The generated
$1/R$ term can drive late-time acceleration without a cosmological constant,
with its scale determined by the entropy deformation parameter rather than
introduced independently. At the same time, its known Dolgov-Kawasaki
instability is reproduced directly by the entropic criterion, while
Eq.~\eqref{eq:scalaronentropy} shows that the corresponding scalaron is
tachyonic. Hence, the thermodynamic and dynamical conclusions are consistent.

The same picture applies to the remaining reconstructions. Barrow and
Tsallis-Cirto entropies with $\delta>1$ generate infrared modifications and
are unstable on this branch. The power-law entanglement entropy generates the
milder correction $R^{1-\beta}$ and satisfies the Dolgov-Kawasaki condition
for $\mu\beta>0$, provided that the additional scalaron condition
$S'(s)>0$ is also fulfilled. The R\'enyi reconstruction is sub-extensive and
provides an infrared modification while satisfying both the Dolgov-Kawasaki
and scalaron stability conditions for arbitrary positive $\lambda$. Whether
it also satisfies the local-gravity and matter-era constraints required for
a fully viable infrared modification is a separate issue, requiring the
matter sector and the analysis of the scalaron on a cosmological background.
This investigation lies beyond the scope of the present work.

\section{Massive branch and horizon charges}
\label{sec:massive}

In this section we turn to the fixed-mass branch and reconstruct the
corresponding $f(R)$ models within the regime of validity of the
area-curvature expansion. We then examine their constant-curvature vacua and
scalaron stability, compare them with the exact maximally symmetric
reconstruction, and finally study the associated weak isolated-horizon
charges.

\subsection{Reconstructed Lagrangians and constant-curvature vacua}
\label{subsec:massive_lagrangians}

We now turn to the branch  \eqref{eq:map_massive}, which is the relevant one 
for a black hole of
fixed mass. As discussed above, the corresponding area-curvature map is known
perturbatively through an expansion around $R=0$. Independently, in the
analysis below we also treat the generalized entropy perturbatively in its
deformation parameter. Throughout this section we use $G=1$ in the
massive-branch expressions, so that $s=\SBH=6\pi c(1+cR)$. Each generalized
entropy is written as
\begin{equation}
S=\SBH\left[1+\epsilon\,\Phi(A)\right]+\mathcal{O}(\epsilon^2) \ ,
\label{eq:pertentropy}
\end{equation}
where $\epsilon$ is the corresponding deformation parameter. Hence, the
Lagrangian can be written as
\begin{equation}
f(R)=R-2\Lambda+\epsilon\,h(R)+\mathcal{O}(\epsilon^2) \ .
\end{equation}

The closed expressions presented below are resummed forms constructed to
reproduce the corresponding perturbative corrections at the retained order
in the deformation parameter. They should not be identified with the exact
finite-deformation quadratures within the massive-branch map, which are
presented in Appendix~\ref{app:massiveexact}.

\paragraph*{R\'enyi case.}
From
$S_{\mathcal R}=\lambda^{-1}\ln(1+\lambda\SBH)
=\SBH-\tfrac{\lambda}{2}\SBH^2+\mathcal{O}(\lambda^2)$
one reads $\Phi_{\mathcal R}=-A/8$, and Eq.~\eqref{eq:master} integrates to
\begin{equation}
f_{\mathcal R}(R)=
\frac{-2c\Lambda+(1+cR)e^{-\frac{3}{2}\pi c\lambda(1+cR)}-1}{c} \ ,
\label{eq:renyimassive}
\end{equation}
whose expansion gives
\begin{equation}
h_{\mathcal R}(R)=-\frac{3\pi}{2}(1+cR)^2 \ .
\label{eq:hrenyi}
\end{equation}
The logarithmic deformation of the entropy translates into a quadratic
curvature correction. We stress that Eq.~\eqref{eq:renyimassive} is a
resummation which is faithful to first order in $\lambda$. The exact
massive-branch R\'enyi reconstruction involves a dilogarithm, and statements
at large $\lambda$ based on Eq.~\eqref{eq:renyimassive} should therefore be
understood accordingly.

\paragraph*{Barrow case.}
From
$S_B=\SBH^{1+\Delta/2}
=\SBH[1+\tfrac{\Delta}{2}\ln\SBH]+\mathcal{O}(\Delta^2)$
one reads $\Phi_B=\tfrac12\ln(A/4)$ and obtains
\begin{equation}
f_B(R)=
\frac{\left[2+\Delta(\ln(6\pi c)-1)\right]
(1+cR)^{1+\frac{\Delta}{2}}}{2c}
-\frac{1+2c\Lambda}{c} \ ,
\label{eq:barrowmassive}
\end{equation}
with
\begin{equation}
h_B(R)=\frac{1+cR}{2c}
\left[\ln\bigl(6\pi c(1+cR)\bigr)-1\right] \ .
\label{eq:hbarrow}
\end{equation}

\paragraph*{Kaniadakis case.}
From
$S_K=\kappa^{-1}\sinh(\kappa\SBH)
=\SBH+\tfrac{\kappa^2}{6}\SBH^3+\mathcal{O}(\kappa^4)$
one reads $\Phi_K=A^2/96$. The absence of a linear term follows from the
symmetry $\kappa\to-\kappa$. We then obtain
\begin{equation}
f_K(R)=
\frac{\sinh\left[2\sqrt3\pi c\kappa(1+cR)\right]}
{2\sqrt3\pi c^2\kappa}+c_1 \ ,
\label{eq:kanimassive}
\end{equation}
with $c_1=-2\Lambda-1/c$ fixed by the Einstein limit and
\begin{equation}
h_K(R)=2\pi^2c(1+cR)^3 \ .
\label{eq:hkani}
\end{equation}
Thus, on this branch the R\'enyi, Barrow and Kaniadakis entropies generate
quadratic, logarithmic and cubic curvature corrections, respectively. The
entropy exponents are preserved rather than reflected, in contrast to the
maximally symmetric branch discussed above.

Having obtained the reconstructed Lagrangians, we proceed to examine their
constant-curvature vacua.

\paragraph*{R\'enyi case.}
Writing $\mathcal{A}=\lambda\pi GM^2$,
Eq.~\eqref{eq:renyimassive} gives
$f=R-2\Lambda-\mathcal{A}(1+cR)^2/c$ and
$f_R=1-2\mathcal{A}(1+cR)$ to first order. The quadratic terms in
Eq.~\eqref{eq:cc} cancel identically, leaving the linear equation
$R(2\mathcal{A}-1)+4\Lambda+2\mathcal{A}/c=0$, with solution
$R=(4\Lambda+2\mathcal{A}/c)/(1-2\mathcal{A})$. Hence,
\begin{equation}
R_0=4\Lambda+8\pi GM^2\Lambda\lambda
+\frac{3\pi}{G}\lambda+\mathcal{O}(\lambda^2) \ .
\label{eq:R0renyi}
\end{equation}

\paragraph*{Barrow case.}
In this case the logarithms prevent Eq.~\eqref{eq:cc} from reducing to a
polynomial, and we solve perturbatively by setting
$R_0=4\Lambda+\Delta R_1$. Evaluating the $\mathcal{O}(\Delta)$ contribution
at $R=4\Lambda$ and defining $X=1+4c\Lambda$, we obtain
\begin{equation}
R_1=2\Lambda\ln(6\pi cX)
-\frac{X}{c}\left[\ln(6\pi cX)-1\right] \ .
\label{eq:R1barrow}
\end{equation}

\paragraph*{Kaniadakis case.}
With $\alpha=4\pi GM^2\kappa/\sqrt3$, Eq.~\eqref{eq:cc} becomes the cubic
\begin{equation}
\frac{\alpha^2c^2}{6}R^3
-\left(1+\frac{\alpha^2}{2}\right)R
+\left(4\Lambda-\frac{\alpha^2}{3c}\right)=0 \ ,
\label{eq:kanicubic}
\end{equation}
whose Cardano discriminant is
\begin{equation}
\Delta_c=\frac{108}{\alpha^6c^6}
\left[
8+\alpha^2(12-144\Lambda^2c^2)
+\alpha^4(24\Lambda c+6)
\right] \ .
\end{equation}
For small $\alpha$ the discriminant is positive, and hence three distinct
real roots exist. The existence of a root continuously connected to the
Einstein branch follows from the implicit function theorem applied to
\begin{equation}
F(R,\alpha)
=-R+4\Lambda+\alpha^2(1+cR)^2
\left(\frac{R}{6}-\frac{1}{3c}\right) \ .
\end{equation}
Since $F(4\Lambda,0)=0$ and
$\pr_RF|_{(4\Lambda,0)}=-1\neq0$, a unique differentiable root
$R(\alpha)$ exists near $\alpha=0$, with $R(0)=4\Lambda$.

In all three cases a real maximally symmetric vacuum exists and is
continuously connected to the Einstein solution. Therefore, each
reconstructed theory admits Schwarzschild-(anti-)de Sitter black-hole
solutions.

\subsection{Scalaron stability and comparison of the two branches}
\label{subsec:scalaroncomparison}

We now evaluate Eq.~\eqref{eq:scalaron} on the backgrounds obtained above.
The results, summarized in Table~\ref{tab:scalaron}, differ significantly
between the models.

\paragraph*{R\'enyi case.}
From Eq.~\eqref{eq:renyimassive}, we obtain
\begin{equation}
m_{\rm sc}^2=
\frac{4+3c\pi\lambda\left[-2+cR_0\left(2-3c\pi(1+cR_0)\lambda\right)\right]}
{9c^2\pi\lambda\left[-4+3c\pi(1+cR_0)\lambda\right]} \ .
\label{eq:msrenyi}
\end{equation}
The denominator vanishes at
$
\lambda_p=\frac{4}{3\pi c(1+cR_0)}$,
which is also the point where $f_{RR}$ changes sign. In particular,
$f_{RR}<0$ for $0<\lambda<\lambda_p$ and $f_{RR}>0$ for
$\lambda>\lambda_p$.

The numerator of Eq.~\eqref{eq:msrenyi} can be written as
\begin{equation}
4+6c\pi(cR_0-1)\lambda
-9c^3\pi^2R_0(1+cR_0)\lambda^2 \ ,
\label{eq:renyinum}
\end{equation}
and is therefore a downward parabola in $\lambda$. Its discriminant is
\begin{equation}
36c^2\pi^2
\left[(cR_0-1)^2+4cR_0(1+cR_0)\right]>0 \ ,
\end{equation}
and the physically relevant positive root is
\begin{equation}
\lambda_0=
\frac{(cR_0-1)+
\sqrt{(cR_0-1)^2+4cR_0(1+cR_0)}}
{3\pi c^2R_0(1+cR_0)} \ .
\label{eq:lambda0}
\end{equation}
For $c>0$ and $R_0>0$ one finds
\begin{equation}
\lambda_0<\lambda_p \ .
\label{eq:lambdaordering}
\end{equation}
Indeed, defining $x=cR_0>0$, one has
\begin{equation}
\frac{\lambda_0}{\lambda_p}
=
\frac{x-1+\sqrt{5x^2+2x+1}}{4x}<1 \ ,
\end{equation}
where the inequality follows from
$(3x+1)^2-(5x^2+2x+1)=4x(x+1)>0$ for $x>0$.

The parameter space therefore separates into three regions.
\begin{itemize}
\item[I.] For $0<\lambda<\lambda_0$, the numerator is positive while the
denominator is negative. Hence $m_{\rm sc}^2<0$, and the scalaron is
tachyonic. Moreover, $f_{RR}<0$.

\item[II.] For $\lambda_0<\lambda<\lambda_p$, both the numerator and the
denominator are negative. Hence $m_{\rm sc}^2>0$, and the scalaron is
non-tachyonic. However, $f_{RR}<0$, and therefore the
Dolgov-Kawasaki condition is violated.

\item[III.] For $\lambda>\lambda_p$, the numerator remains negative while
the denominator becomes positive. Hence $m_{\rm sc}^2<0$, although
$f_{RR}>0$.
\end{itemize}
Thus, the scalaron is tachyon-free only in the finite interval
\begin{equation}
\lambda_0<\lambda<\lambda_p \ ,
\label{eq:renyiwindow}
\end{equation}
but this interval lies entirely in the region where $f_{RR}<0$. Therefore,
there is no positive-$\lambda$ interval for which the resummed R\'enyi
reconstruction simultaneously satisfies the Dolgov-Kawasaki condition and
the non-tachyonic scalaron condition.

This result is consistent with the branch-independent relation
\eqref{eq:lemma_stab}. Since the R\'enyi entropy is sub-extensive, while the
massive branch requires super-extensivity, the exact massive-branch
reconstruction must have $f_{RR}<0$. The sign change at $\lambda_p$ in the
resummed Lagrangian \eqref{eq:renyimassive} occurs outside the regime in
which the first-order reconstruction is controlled and should not be
interpreted as a physical recovery of stability. The exact quadrature of
Appendix~\ref{app:massiveexact} confirms that the R\'enyi reconstruction is
Dolgov-Kawasaki unstable for every $\lambda>0$.

\paragraph*{Barrow case.}
Equation~\eqref{eq:barrowmassive} gives
\begin{equation}
m_{\rm sc}^2=\frac{2+cR_0(2-\Delta)}{3c\Delta} \ ,
\label{eq:msbarrow}
\end{equation}
whose numerator and denominator are positive for $c>0$, $R_0>0$ and
$0<\Delta\le1$. The reconstructed Barrow theory is therefore stable over
its entire physical parameter range, without any additional constraint on
$\Delta$. As $\Delta\to0^+$,
\begin{equation}
m_{\rm sc}^2\sim
\frac{2(1+cR_0)}{3c\Delta}
\longrightarrow+\infty \ ,
\end{equation}
and thus the scalaron becomes infinitely massive and decouples. At the same
time, the Barrow entropy continuously returns to $\SBH$ and the theory
returns to General Relativity, where the extra scalar mode is absent. At
maximal deformation $\Delta=1$, one still has
\begin{equation}
m_{\rm sc}^2=\frac{2+cR_0}{3c}>0 \ ,
\end{equation}
and hence even the strongest admissible fractal deformation leaves the
constant-curvature vacuum tachyon-free. Among the three models, this is the
most robust behavior on the massive branch.

\paragraph*{Kaniadakis case.}
Equation~\eqref{eq:kanimassive} gives
\begin{equation}
m_{\rm sc}^2=
\frac{\sqrt3}{18c^2\pi\kappa}
\coth\left(2\sqrt3\pi c(1+cR_0)\kappa\right)
-\frac{R_0}{3} \ ,
\label{eq:mskani}
\end{equation}
whose sign cannot be determined in closed form. Expanding consistently in
the perturbative regime $\kappa\ll1$, we obtain
\begin{align}
m_{\rm sc}^2=\;&
\frac{1}{36c^3\pi^2\kappa^2(1+cR_0)}
-\frac{2cR_0-1}{9c}\notag\\
&-\frac{4\pi^2c}{45}\kappa^2(1+cR_0)^3
+\mathcal{O}(\kappa^4) \ .
\label{eq:mskaniseries}
\end{align}
There is no linear term, in agreement with the symmetry
$\kappa\to-\kappa$ of the Kaniadakis entropy. The leading contribution
$\sim[36\pi^2c^3(1+cR_0)\kappa^2]^{-1}>0$ diverges as $\kappa\to0$,
while the remaining terms remain finite. Therefore, the reconstructed
Kaniadakis theory is free of tachyonic instabilities for sufficiently small
$\kappa$. Moreover, Eq.~\eqref{eq:lemma_stab} implies $f_{RR}>0$ on the
massive branch, since the Kaniadakis entropy is super-extensive. Beyond the
perturbative regime, the competition between the hyperbolic cotangent and
the $-R_0/3$ term may drive $m_{\rm sc}^2$ through zero, but this regime lies
outside the validity of the perturbative reconstruction.

\begin{table}[b]
\centering
\begin{ruledtabular}
\begin{tabular}{lccc}
Entropy & $m_{\rm sc}^2>0$ & DK condition & Verdict \\
\hline
R\'enyi &
$\lambda_0<\lambda<\lambda_p$ &
$f_{RR}<0$ &no viable window \\
&& in this interval &
\\
Barrow &
$0<\Delta\le1$ &
$f_{RR}>0$ &
stable \\
Kaniadakis &
$\kappa\ll1$ &
$f_{RR}>0$ &$\!$
perturbatively stable \\
\end{tabular}
\end{ruledtabular}
\caption{Scalaron and Dolgov-Kawasaki stability of the massive-branch
reconstructions. For the R\'enyi case the scalaron is non-tachyonic only for
$\lambda_0<\lambda<\lambda_p$, but $f_{RR}<0$ throughout this interval, and
therefore no fully viable stability window exists. The Barrow and Kaniadakis
results agree with the super-extensivity criterion
\eqref{eq:lemma_stab}.}
\label{tab:scalaron}
\end{table}

The above results allow us to compare directly the two reconstruction
branches. The two sets of Lagrangians are not approximations of one another.
They correspond to different curves in the space of constant-curvature vacua
and answer different physical questions. The massive branch determines the
Lagrangian that reproduces a given entropy for a black hole of fixed mass,
leading to corrections organized in powers of $1+cR$ and retaining a
dependence on $M$. On the other hand, the maximally symmetric branch
determines the Lagrangian associated with a cosmological horizon, leading to
corrections organized in powers of $R$ and containing no free parameter
beyond $\sigma=12\pi/G$.

The main structural difference between the two branches lies in the exponent
mapping. On the massive branch $s\propto1+cR$, and a term $a\,s^q$ in the
entropy generates a term proportional to $(1+cR)^q$ in the Lagrangian.
Hence, the exponent is preserved. On the maximally symmetric branch
$s\propto R^{-1}$, and the same entropy term generates $R^{2-q}$, so the
exponent is reflected about unity. For instance, the Barrow entropy generates
a correction proportional to $(1+cR)^{1+\Delta/2}$ on the
massive branch and an infrared correction $R^{1-\Delta/2}$ on the maximally
symmetric branch. These different forms arise because a different horizon is
used in each reconstruction.

The Dolgov-Kawasaki stability properties are correspondingly inverted. Since
the sign of $ds/dR$ differs between the two branches,
Eq.~\eqref{eq:lemma_stab} implies that the same monotonicity property of
$S(s)/s$ gives opposite conditions. Thus, the Barrow reconstruction is stable
on the massive branch and unstable on the maximally symmetric one. The
Kaniadakis reconstruction is perturbatively stable on the massive branch and
unstable on the maximally symmetric one. Conversely, the R\'enyi
reconstruction is Dolgov-Kawasaki stable on the maximally symmetric branch
but unstable on the massive branch. The explicit scalaron analysis above is
consistent with this result, since the only interval where the resummed
R\'enyi scalaron is non-tachyonic lies entirely in the region $f_{RR}<0$.
Hence, there is no contradiction between the two branches, since they lead
to different reconstructed Lagrangians.

The comparison shows that the entropy-Lagrangian correspondence depends on
the branch on which the reconstruction is performed. Therefore, physical
conclusions concerning the resulting gravitational theory should always be
associated with the corresponding horizon branch. For cosmological
applications, where the relevant horizon is cosmological, the exact
maximally symmetric reconstruction is the natural choice. On the other hand,
for the thermodynamics of an individual black hole, the massive branch is
the appropriate one.

\subsection{First-order form and horizon charges}
\label{sec:firstorder}

We finally verify, within the covariant phase space of weak isolated horizons,
that the conserved charge of the reconstructed theory reproduces the entropy
from which it was constructed. This provides an independent consistency check.
The Wald entropy is defined at a Killing horizon of a stationary spacetime,
whereas the isolated-horizon charge is quasi-local and does not require the
existence of a global Killing field. The scalar-tensor representation is 
particularly useful here, since it makes
the additional scalar degree of freedom explicit and provides the natural
first-order formulation for the subsequent horizon-charge analysis.

Metric $f(R)$ gravity is dynamically equivalent to a non-minimally coupled
scalar-tensor theory \cite{DeFelice:2010aj}.
Introducing an auxiliary field $\chi$, we write
$\mathcal{A}[g,\chi]=(16\pi G)^{-1}\int\sqrt{-g}
[f(\chi)-f_\chi(\chi)(\chi-R)]$, where $f_\chi=\partial_\chi f$.
Variation with respect to $\chi$ enforces $\chi=R$ whenever
$f_{\chi\chi}\neq0$. Defining
\begin{equation}
\phi\equiv f_R(\chi) \ , \qquad
V(\phi)=\phi\chi(\phi)-f(\chi(\phi)) \ ,
\label{eq:legendre}
\end{equation}
the action takes the Jordan-frame form
$\mathcal{A}=(16\pi G)^{-1}\int\sqrt{-g}\,[\phi R-V(\phi)]$.
In first-order variables it becomes
\begin{align}
16\pi G\,\mathcal{A}(e,A)=&
\int_M\phi\,\Sigma_{IJ}\wedge F^{IJ}\notag\\
&-\frac{1}{4!}\int_MV(\phi)\epsilon_{IJKL}
e^I\wedge e^J\wedge e^K\wedge e^L \ ,
\label{eq:firstorderaction}
\end{align}
where $\Sigma_{IJ}=\tfrac12\epsilon_{IJKL}e^K\wedge e^L$ and $F^{IJ}$ is
the curvature of the Lorentz connection $A^{IJ}$. Since all backgrounds
considered here have $R=R_0$, the field $\phi=f_R(R_0)$ is constant.
Therefore, the connection equation $D(\phi\Sigma_{IJ})=0$ reduces to
$D\Sigma_{IJ}=0$, and no conformal rescaling of the tetrad is required
\cite{Ashtekar:2003jh}.

We recall the relevant definitions
\cite{Ashtekar:1998sp,Ashtekar:2000hw,Ashtekar:1999yj,Ashtekar:2001is,
Ashtekar:2004cn}. A co-dimension-one null surface $\Delta$ equipped with a
degenerate metric $q_{ab}$ of signature $(0,+,+)$ and an equivalence class of
null normals $[C\ell^a]$, with $C$ a positive constant, is a
\emph{non-expanding horizon} (NEH) if (i) it is topologically
$S^2\times\mathbb{R}$, (ii) its null normals are expansion free,
$\theta_{(\ell)}\hat{=}0$, (iii) $-T^a{}_b\ell^b$ is a future-directed causal
vector field, and (iv) the field equations hold on $\Delta$. An NEH is a
\emph{weak isolated horizon} (WIH) if it admits an equivalence class for which
the connection on the normal bundle is Lie dragged along the generators,
$\pounds_\ell\omega_{(\ell)}\hat{=}0$. This characterization is quasi-local,
since it describes a black hole in local equilibrium without requiring the
exterior spacetime to be stationary or the existence of a global Killing
field. The WIH boundary conditions break the bulk $SO(3,1)$ symmetry to the
subgroup preserving $[\ell^a]$, while boosts in the $(\ell,n)$ plane remain
genuine horizon symmetries \cite{Basu:2010hv}.

The on-shell variation of Eq.~\eqref{eq:firstorderaction} is a pure boundary
term,
$16\pi G\,\delta\mathcal{A}
=\int_{\pr\mathcal{M}}\phi\,\Sigma_{IJ}\wedge\delta A^{IJ}$,
from which the symplectic potential and current follow as
\begin{align}
16\pi G\,\theta(\delta)
&=\phi\,\Sigma_{IJ}\wedge\delta A^{IJ} \ ,\\
16\pi G\,J(\delta_1,\delta_2)
&=\delta_1(\phi\Sigma_{IJ})\wedge\delta_2A^{IJ}
-(1\leftrightarrow2) \ .
\end{align}
For a manifold without boundary, the closure of $J$ guarantees conservation
of the symplectic structure. An inner boundary modifies this result.
However, since a WIH represents equilibrium, it must not act as a genuine
flux source, and the pullback of $J$ to $\Delta$ is exact, $J=dj$. Therefore,
the horizon contribution reduces to the initial and final cross-sections, and
the conserved symplectic structure acquires the boundary term
\begin{multline}
16\pi G\,\Omega(\delta_1,\delta_2)
=\int_M\delta_1(\phi\Sigma_{IJ})\wedge\delta_2A^{IJ}
-(1\leftrightarrow2)\\
-\oint_{S_\Delta}
\delta_1(\phi\,{}^2\epsilon)\delta_2\psi
-\delta_2(\phi\,{}^2\epsilon)\delta_1\psi \ ,
\end{multline}
where ${}^2\epsilon$ is the volume form on the horizon cross sections and $\pounds_\ell\psi\hat{=}\kappa_{(\ell)}$, with $\psi=0$ on the initial
cross section of the horizon. The Hamiltonian function generating the residual boost labelled by the parameter $\eta$ is then
\cite{Chatterjee:2020iuf,Devdutt:2026pia,Garg:2026toh}:
\begin{equation}
H_{\eta}=\mathcal{N}\oint_{S_\Delta}\phi\,{}^2\epsilon
=\mathcal{N} f_R(R_0)A_\Delta \ ,
\label{eq:boost}
\end{equation}
where $\mathcal{N}$ is a normalization constant fixed by the Einstein limit,
namely $\mathcal{N}=1/(4G)$, or $\mathcal{N}=1/4$ in units $G=1$.

We now apply this construction to the three massive-branch reconstructions.

\paragraph*{R\'enyi case.}
From Eq.~\eqref{eq:renyimassive},
\begin{equation}
\phi(R)=\tfrac12
e^{-\frac32\pi c\lambda(1+cR)}
\left[2-3\pi c\lambda(1+cR)\right] \ ,
\end{equation}
which can be inverted through the Lambert function as
$R(\phi)=[2-3\pi c\lambda-2W(e\phi)]/(3\pi c^2\lambda)$.
The Legendre transform \eqref{eq:legendre} then gives
\begin{equation}
V(\phi)=
\frac{3\pi c\lambda(2c\Lambda-\phi+1)-2\phi W(e\phi)
-\dfrac{2\phi}{W(e\phi)}+4\phi}
{3\pi c^2\lambda} \ .
\end{equation}

\paragraph*{Barrow case.}
From Eq.~\eqref{eq:barrowmassive},
\begin{equation}
\phi(R)=\frac{\Delta+2}{4}
\left(\Delta\ln(6\pi c)-\Delta+2\right)
(1+cR)^{\Delta/2} \ ,
\end{equation}
which can be inverted algebraically, yielding
\begin{align}
V(\phi)=\frac{1}{c}\Bigg[
&1+2c\Lambda-\phi
+\frac{\Delta\,16^{1/\Delta}\phi}{\Delta+2}\notag\\
&\times\left(
\frac{\phi}
{(\Delta+2)(2-\Delta+\Delta\ln(6\pi c))}
\right)^{2/\Delta}
\Bigg] \ .
\end{align}

\paragraph*{Kaniadakis case.}
From Eq.~\eqref{eq:kanimassive},
$\phi(R)=\cosh[2\sqrt3\pi c\kappa(1+cR)]$, and hence
\begin{equation}
R(\phi)=
\frac{\cosh^{-1}\phi-2\sqrt3\pi c\kappa}
{2\sqrt3\pi c^2\kappa} \ .
\end{equation}
The corresponding potential is
\begin{align}
V(\phi)=\frac{1}{6\pi c^2\kappa}\Big[
&6\pi c\kappa(2c\Lambda-\phi+1)\notag\\
&-\sqrt{3(\phi^2-1)}
+\sqrt3\,\phi\cosh^{-1}\phi
\Big] \ .
\end{align}

Substituting each $\phi=f_R(R_0)$ into Eq.~\eqref{eq:boost} and working in
units $G=1$, such that $\mathcal{N}=1/4$, we obtain
\begin{align}
H_\eta^{(\mathcal R)}
&=\frac{A_\Delta}{4}
\left[
1-\frac{\lambda A_\Delta}{8}
+\mathcal{O}(\lambda^2)
\right] \ ,\\
H_\eta^{(B)}
&=\frac{A_\Delta}{4}
\left[
1+\frac{\Delta}{2}\ln\left(\frac{A_\Delta}{4}\right)
+\mathcal{O}(\Delta^2)
\right] \ ,\\
H_\eta^{(K)}
&=\frac{A_\Delta}{4}
\left[
1+\frac{\kappa^2A_\Delta^2}{96}
+\mathcal{O}(\kappa^4)
\right] \ .
\end{align}
These expressions coincide term by term with the perturbative R\'enyi,
Barrow and Kaniadakis entropies. Hence, the generalized entropy is recovered
not only as a thermodynamic input, but also as a conserved geometric charge
associated with a horizon symmetry. Moreover, the same normalization
$\mathcal{N}=1/4$ applies in all three cases, without introducing any
model-dependent rescaling.

\section{Conclusions}
\label{sec:conclusions}

Generalized horizon entropies provide a natural way to incorporate possible
departures from the Bekenstein-Hawking area law and have been widely used in
black-hole thermodynamics and cosmology. However, if the entropy of a horizon
is modified, it is important to determine whether this modification can be
associated with an underlying gravitational dynamics. In the present work we
have addressed this inverse problem within metric $f(R)$ gravity, using the
Wald entropy relation to reconstruct the gravitational Lagrangian from a
prescribed entropy-area functional.

A central point of the analysis is that the entropy functional alone does not
uniquely determine the Lagrangian. The Wald relation fixes $f_R$ as a function
of the horizon area, while the gravitational action requires $f_R$ as a
function of the curvature. Hence, an area-curvature relation $A(R)$ must first
be specified. For Schwarzschild-de Sitter geometries the horizon area is not
uniquely determined by the curvature, since the mass provides an additional
parameter. Therefore, the reconstruction is defined only after selecting a
one-parameter branch in the space of constant-curvature geometries. We have
considered two physically distinct choices, namely the fixed-mass massive
branch and the maximally symmetric branch.

The maximally symmetric branch is particularly useful, since its cosmological
horizon obeys the exact and mass-independent relation $A=48\pi/R$. As a
result, the reconstruction reduces to the single quadrature
\eqref{eq:exactmaster}, without any expansion in the entropy deformation
parameter. We have applied this relation to the Bekenstein-Hawking,
Tsallis-Cirto, Barrow, R\'enyi, Kaniadakis, logarithmically corrected and
power-law entanglement entropies. The corresponding $f(R)$ Lagrangians are
obtained in closed form and are summarized in Table~\ref{tab:exact}. In
particular, the Bekenstein-Hawking entropy reproduces Einstein gravity
identically, Tsallis-Cirto and Barrow entropies lead to power-law
Lagrangians, R\'enyi entropy gives an elementary logarithmic structure,
Kaniadakis entropy leads to a hyperbolic sine integral, the logarithmic
entropy produces an $R^2\ln R$ correction, and the power-law entanglement
entropy generates an $R^{1-\beta}$ modification.

The exact construction also allows for several general results that go beyond
the individual entropy models. An entropy contribution proportional to
$\SBH^q$ generates a curvature contribution proportional to $R^{2-q}$, with
the exceptional case $q=2$ giving a logarithmic term. Thus, on the maximally
symmetric branch the entropy exponent is reflected about the extensive value
$q=1$. This mapping determines the curvature regime in which a given
correction becomes important. In particular, entropy terms with $q>1$
generate corrections with curvature power smaller than unity and therefore
have an infrared character. A characteristic example is provided by the
Kaniadakis entropy, whose leading correction generates the $1/R$ model of
Carroll, Duvvuri, Trodden and Turner \cite{Carroll:2003wy}, with its scale
fixed directly by the entropy deformation parameter. On the other hand, a
logarithmic entropy correction generates the $R^2\ln R$ structure associated
with higher-curvature effective descriptions.

A second general result concerns the stability of the reconstructed theories.
We have derived the branch-independent relation
$
f_{RR}=\frac{ds}{dR}\frac{d}{ds}\left(\frac{S}{s}\right),$
which converts the Dolgov-Kawasaki condition into a simple monotonicity
criterion for $S(s)/s$. The sign of $ds/dR$ is opposite on the two branches,
and therefore the corresponding stability conditions are reversed. On the
maximally symmetric branch Dolgov-Kawasaki stability requires
$d(S/s)/ds<0$, while on the massive branch it requires
$d(S/s)/ds>0$. Moreover, on the maximally symmetric branch we obtained the
additional relation
$
m_{\rm sc}^2=\frac{S'(s)}{3f_{RR}} $, 
which directly connects the scalaron mass to the entropy functional.
Consequently, for a positive and monotonically increasing entropy,
ghost freedom, Dolgov-Kawasaki stability and the absence of a tachyonic
scalaron can all be formulated through simple properties of the entropy
itself.

We have additionally studied the massive branch, which is relevant to a black
hole of fixed mass. In this case the area-curvature relation is obtained from
an expansion around the Schwarzschild solution, independently of the
perturbative expansion in the entropy deformation parameter used in the main
analysis. We reconstructed the R\'enyi, Barrow and Kaniadakis Lagrangians,
determined the corresponding constant-curvature vacua and examined their
scalaron stability. The Barrow reconstruction is stable throughout its
physical parameter range, while the Kaniadakis reconstruction is
perturbatively stable for sufficiently small deformation. The R\'enyi case
provides an instructive contrast. Although the resummed expression possesses
a finite interval in which the scalaron is non-tachyonic, this interval lies
entirely in the region $f_{RR}<0$. Hence, no positive-$\lambda$ interval
simultaneously satisfies the Dolgov-Kawasaki and non-tachyonic scalaron
conditions. This agrees with the branch-independent criterion and with the
exact massive-branch quadrature presented in Appendix~\ref{app:massiveexact}.

The comparison of the two branches shows that they should not be regarded as
different approximations to the same reconstructed theory. They correspond
to different curves in the space of constant-curvature geometries and thus
lead to different Lagrangians. On the massive branch entropy powers are
preserved through the dependence on $1+cR$, while on the maximally symmetric
branch they are reflected according to $q\rightarrow2-q$. Their stability
properties are correspondingly different. This branch dependence is therefore
not a technical detail but an intrinsic part of the entropy-Lagrangian
correspondence.

Finally, we have examined the reconstructed theories in first-order form and
within the covariant phase space of weak isolated horizons. The conserved
charge associated with the residual horizon boost reproduces, with the same
Einstein normalization, the generalized entropy from which each theory was
constructed. This provides a quasi-local consistency check that is
independent of the stationary Killing-horizon formulation used in the Wald
reconstruction and further supports the thermodynamic interpretation of the
resulting gravitational theories.

There are several directions in which the present framework can be extended.
The most immediate one is the inclusion of matter and the study of the
scalaron on cosmological and astrophysical backgrounds, which is necessary
for a complete assessment of local-gravity and matter-era stability
constraints. It would also be interesting to investigate the cosmological
dynamics of the exact Lagrangians obtained here, especially those containing
infrared corrections that may affect late-time acceleration. On the
gravitational side, the reconstruction can be generalized to charged and
rotating horizons, other exact branches such as Nariai-type geometries,
higher-dimensional spacetimes and higher-curvature theories, including
Lovelock gravity. Such extensions may clarify to what extent the
entropy-Lagrangian correspondence found here represents a more general
connection between modified horizon thermodynamics and gravitational
dynamics.

\begin{acknowledgments}
Ankit Anand is financially supported by the Institute's postdoctoral fellowship 
at IIT Kanpur. Sahil Devdutt acknowledges financial support from DST under 
Grant No.\ DST/INSPIRE Fellowship/2020/IF200537. SD also acknowledges helpful initial discussions with Sk Jahanur Hoque and Antariksha Basu. The authors acknowledge the 
contribution of COST Actions CA21106 ``COSMIC WISPers in the Dark Universe: 
Theory, astrophysics and experiments'', CA21136 ``Addressing observational 
tensions in cosmology with systematics and fundamental physics (CosmoVerse)'', 
CA23130 ``Bridging high and low energies in search of quantum gravity 
(BridgeQG)'', and CA24101 ``Testing Fundamental Physics with Seismology''.  
\end{acknowledgments}


\appendix

\section{Exact quadratures on the massive branch}
\label{app:massiveexact}

The perturbative character of the massive-branch analysis in
Sec.~\ref{subsec:massive_lagrangians} originates from the area-curvature map
\eqref{eq:map_massive}, which is obtained as a first-order expansion of the
horizon cubic around the Schwarzschild point. It does not originate from the
integration of the master relation itself. In fact, within the adopted
massive-branch map the quadrature can be performed exactly in the entropy
deformation parameter. Writing $s=s_0(1+cR)$, with $s_0=A_0/4G$, such that
$ds=s_0c\,dR$, Eq.~\eqref{eq:master} integrates to
\begin{equation}
f(R)=\frac{1}{s_0c}
\int^{\,s_0(1+cR)}\frac{S(s')}{s'}\,ds'+c_1 \ .
\label{eq:massivemaster}
\end{equation}
For the three entropies considered in the main text this gives
\begin{align}
f_B(R)&=\frac{2}{s_0c\,(2+\Delta)}
\,s^{1+\frac{\Delta}{2}}+c_1 \ ,
\label{eq:barrowappendix}\\[3pt]
f_K(R)&=\frac{1}{s_0c\,\kappa}\,
\Shi(\kappa s)+c_1 \ ,
\label{eq:kaniappendix}\\[3pt]
f_{\mathcal R}(R)&=-\frac{1}{s_0c\,\lambda}\,
\mathrm{Li}_2(-\lambda s)+c_1 \ ,
\label{eq:renyiappendix}
\end{align}
where $\mathrm{Li}_2$ is the dilogarithm,
$\mathrm{Li}_2(z)=-\int_0^z\ln(1-t)t^{-1}dt$, and $c_1$ is fixed so that
the undeformed limit reproduces $f(R)=R-2\Lambda$. These expressions involve
no expansion in $\Delta$, $\kappa$, or $\lambda$ and have been verified by
direct differentiation against Eq.~\eqref{eq:master}.

A first observation concerns the exponent mapping. Equations
\eqref{eq:barrowappendix}-\eqref{eq:renyiappendix} confirm directly that the
massive branch preserves the entropy exponents. In particular, the Barrow
entropy produces a pure power of $s\propto1+cR$, while on the maximally
symmetric branch the same entropy generates the reflected power
$R^{1-\Delta/2}$ according to Eq.~\eqref{eq:lemma}.

A second observation concerns the relation between these exact quadratures and
the closed expressions used in the main text. Equations
\eqref{eq:barrowmassive}, \eqref{eq:kanimassive}, and
\eqref{eq:renyimassive} are resummed forms constructed from the corresponding
leading deformation expansions. They reproduce
Eqs.~\eqref{eq:barrowappendix}-\eqref{eq:renyiappendix} at the orders retained
in the perturbative treatment, namely $\mathcal{O}(\Delta)$,
$\mathcal{O}(\kappa^2)$, and $\mathcal{O}(\lambda)$, respectively, but they
should not be identified with the exact quadratures at finite deformation.
In particular, the exact R\'enyi reconstruction is dilogarithmic and the exact
Kaniadakis reconstruction involves the hyperbolic sine integral. Thus, the
expressions used in the main text are the appropriate ones for the
perturbative analysis, while Eqs.~\eqref{eq:barrowappendix}-
\eqref{eq:renyiappendix} display the full dependence on the deformation
parameters within the approximate map \eqref{eq:map_massive}.

Finally, the exact quadratures provide an independent check of the
branch-independent stability relation \eqref{eq:lemma_stab}. Differentiating
Eqs.~\eqref{eq:barrowappendix}-\eqref{eq:renyiappendix} twice gives
\begin{align}
\pr^2_Rf_B&=\frac{\Delta}{2}\,s_0c\,
s^{\frac{\Delta}{2}-1}>0 \ ,
\label{eq:appfB}\\[3pt]
\pr^2_Rf_K&=s_0c\,
\frac{\kappa s\cosh(\kappa s)-\sinh(\kappa s)}
{\kappa s^{2}}>0 \ ,
\label{eq:appfK}\\[3pt]
\pr^2_Rf_{\mathcal R}&=s_0c\,
\frac{1}{\lambda s^{2}}
\left[
\frac{\lambda s}{1+\lambda s}
-\ln(1+\lambda s)
\right]<0 \ ,
\label{eq:appfR}
\end{align}
where the signs follow from
$x\cosh x>\sinh x$ and $x/(1+x)<\ln(1+x)$ for $x>0$. Therefore, within the
massive-branch map, the super-extensive Barrow and Kaniadakis entropies satisfy
the Dolgov-Kawasaki condition for all values of their deformation parameters,
whereas the sub-extensive R\'enyi entropy violates it for every $\lambda>0$.
This is precisely the inverted stability pattern predicted by
Eq.~\eqref{eq:lemma_stab}, now obtained without expanding in the entropy
deformation parameters.

This result also clarifies the R\'enyi stability window found in
Sec.~\ref{subsec:scalaroncomparison}. The apparent recovery of positivity for
$\lambda>\lambda_p$ arises from the resummed Lagrangian
\eqref{eq:renyimassive} and is not a property of the exact R\'enyi quadrature
within the adopted massive-branch map.

\end{document}